\documentclass[runningheads]{llncs}
\usepackage{graphicx}
\usepackage{url}

\usepackage{latexsym}

\def\Underline{\setbox0\hbox\bgroup\let\\\endUnderline}
\def\endUnderline{\vphantom{y}\egroup\smash{\underline{\box0}}\\}
\def\|{\verb|}

\begin{document}

\title{Linking Contexts from Distinct Data Sources \\in Zero Trust Federation}

\author{Masato Hirai \and
	Daisuke Kotani \and
	Yasuo Okabe}

\institute{Kyoto University, Yoshida-honmachi, Sakyo-ku, Kyoto 606-8501 JAPAN}
\maketitle
\begin{abstract}
	An access control model called Zero Trust Architecture (ZTA) has attracted attention.
	ZTA uses information of users and devices, called context, for authentication and authorization.
	Zero Trust Federation (ZTF) has been proposed as a framework for extending an idea of identity federation to support ZTA.
	ZTF defines CAP as the entity that collects context and provides it to each organization (Relying Party; RP) that needs context for authorization based on ZTA.
	To improve the quality of authorization, CAPs need to collect context from various data sources.
	However, ZTF did not provide a method for collecting context from data sources other than RP.
	In this research, as a general model for collecting context in ZTF, we propose a method of linking identifiers between the data source and CAP.
	This method provides a way to collect context from some of such data sources in ZTF.
	Then, we implemented our method using RADIUS and MDM as data sources and confirmed that their contexts could be collected and used.

\end{abstract}

\section{Introduction}\label{sec-intro}	
In Zero Trust Architecture (ZTA)\cite{nist800-207}, an organization always verifies the origin of access using data called context from various sources and authorizes access using verification results.
Context includes continuously changing information such as the location of the user or device, access history, as well as the surrounding conditions of the accessing source.

To extend ZTA for Identity Federation(IdF), Zero Trust Federation (ZTF) \cite{hatakeyama-ztf} is proposed as a model to extend ZTA to make authorization decisions using the context of multiple organizations in IdF.
The ZTF introduces Context Attribute Provider (CAP) as the entity to share context in the federation.
CAP collects context independently of the organization and provides context to each organization (Relying Party; RP) with the user's authorization.
However, the method of CAPs collecting context from sources other than RPs was not described in ZTF.
To make more precise authorization decisions in ZTF, it needs to collect diverse contexts from more data sources, including those other than RPs.

The data sources that can provide context are diverse.
For the data source to work as CAP, a huge amount of additional implementations are required, including communication with an authorization server to obtain the user's authorization status and sending the context to the RP.
These implementations are not always possible for some data sources.
Designing for each case would increase the cost of implementation, making it difficult to collect context from various data sources.

This study discusses the methods of collecting context to show the desirable architecture as a CAP, and proposes design for data sources that are difficult to implement additionally, which requires particular consideration.

It was unclear that the relationship between data source and CAP because CAP was proposed to collect and provide context in \cite{hatakeyama-ztf}.
Therefore, we transfer CAP's role of collecting context to data sources and define the data sources as Context Collector (CtxC).
This leaves the CAP's role only to provide, and what we should discuss is how to send context from CtxC to the CAP.
Contexts collected by CtxC usually contain the CtxC's unique user or device identifier (CtxC-id).
For the CAP to provide the context received from CtxC to each RP, the CAP must be able to determine to which user or device in CAP the context should be mapped.
Thus, the problem is how to map the CtxC-id in the context to the CAP-id in the CAP.
This study refers to this as linking context.

In this study, we propose a design for CtxC to perform the linking context in three cases:
(1) a case where CtxC is easy to be extended for pseudonymous ID sharing with CAP, which was proposed in the previous ZTF;
(2) a case where the administrator of CtxC and CAP is the same, which does not require additional implementation but trusts the administrators of both CtxC and CAP;
(3) a case where CtxC uses certificates to authenticate devices or users, which does not require to trust the administrator links contexts properly.
In (1), the CtxC and CAP share a pseudonymous ID for each user or device in advance, and the CtxC includes the pseudonymous ID in the context before sending it to the CAP so that the CAP can link the context to CAP-id.
In (2), the administrator confirms the correspondence between identifiers in CtxC and CAP directly and places the correspondence table of the identifiers in CAP.
In (3), CAP requests the certificate used in CtxC from the user or device and links identifiers which are related to the same certificates.

Furthermore, as a specific design for case (3), this study presents implementations for authentication and authorization by verifying factors such as which LAN the device is connected to and whether the device's OS has been compromised.
As in CtxC, one implementation uses a RADIUS server using EAP-TLS and the other uses MDM.
Through these implementations, we show the way of linking contexts using certificates.

\section{Related Research}\label{sec-background}
ZTA\cite{nist800-207} is a new access control model in contrast to the traditional access control method, the perimeter model.
ZTA controls access by constantly verifying access requests.
To verify access, ZTA uses context, which includes static information as well as dynamic information such as the situation surrounding the user or device.

ZTF\cite{hatakeyama-ztf} is a framework for RPs inside IdFs to federate the context to authenticate and authorize users like ZTA.
ZTA typically has a single organization centrally collecting context\cite{kindervag2010build}.
However, in IdF, contexts are dispersed across multiple RPs and cannot be aggregated.

Therefore, ZTF defines Context Attribute Provider (CAP), which collects and provides context across IdFs.
CAP is proposed as a framework that enables authentication and authorization using sufficient types and amounts of context, even for IdFs used infrequently.

As a method for CAP to provide context to the RP, ZTF proposes using CAEP\cite{caep} for continuous authentication and authorization and UMA\cite{uma-fed} to authorize access under user's authorization.

However, it was unclear in the ZTF how to collect context from sources other than RPs.
Collecting various contexts from sources including non-RP ones will improve authorization quality and will be required.
For example, by using records of entering and leaving a room as context, it is possible to know where users are.
On the other hand, the system of collecting such records does not provide a way of authenticating users directly.
If the system is designed with different authentication requirements, it is too difficult to implement such authenticate features.
Therefore, the same protocols cannot be used as RPs when sharing a context with the RP.

\section{The Method of Linking Context}\label{sec-cap}
\subsection{Definition of Context Collector(CtxC)}\label{subsec-def-cap}
Context is collected by RPs and CAPs in ZTF, but there are important data sources that belong neither to RP nor CAP.
In order for the data source to be a CAP, the data source must manage the context based on the user's authorization to provide the context to the RP.
However, these features are not suitable for all data sources, such as embedded systems.
Therefore, to pursue a more desirable architecture, we define a data source as a Context Collector (CtxC).
This means that we separate CAP's roles into collection and provision of context, and transfer the collection role to CtxC.
In other words, CtxC is responsible for collecting the context, and CAP is responsible for verifying the user's authorization and providing the context to the RP.
The CAP is redefined below as always collecting context indirectly from the CtxC.

\begin{figure}[htb]
	\centering
	\includegraphics[width=9cm]{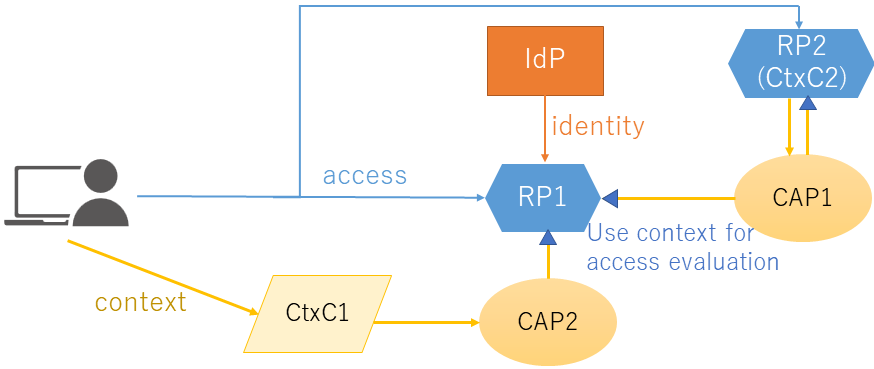}
	\caption{ZTF organized by defining CtxC}\label{fig:ztf-ctxc}
\end{figure}

This change in ZTF is illustrated in Fig. \ref{fig:ztf-ctxc}.
CAP2 receives the context indirectly from CtxC1.
Also, RP2 is regarded as CtxC because it collects context directly from the user.
A single CAP handles one or more CtxCs.

\textbf{The Reason Why Linking Context Is Necessary}

When CtxC collects contexts, the contexts are usually identified with CtxC's user/device identifier (CtxC-id).
In contrast to CtxC-id, we call the identifier of the user in CAP as CAP-id.
CAP must check the mapping between CtxC-id and CAP-id to manage user's authorization of RPs.
If CAP obtains that mapping, it can manage authorization status using such protocols as OAuth2.0 and UMA.
Thus, to provide the context for RPs properly, CAP needs a method of mapping between the CtxC-id and the CAP-id.
Therefore, we call this mapping as linking context, and discuss the method of linking context.
Since it is difficult to link context without making any assumptions about CtxC or CtxC-id, we considered three cases along with use cases.
We propose solutions for each of the cases below.
One is the case where CtxC is easily extensible to share IDs using some protocols as in web applications.
Another is the case where CtxC and the CAP administrator are the same.
This case will be applicable for the entry/exit record system in a company.
The other is the case where CtxC authenticates with client certificates.
This case can apply to Radius server using EAP-TLS\cite{eaptls-rfc5216}.

\subsection{Linking context}\label{subsec-ctxc-easy}
\begin{figure}[htb]
	\centering
	\includegraphics[width=6cm]{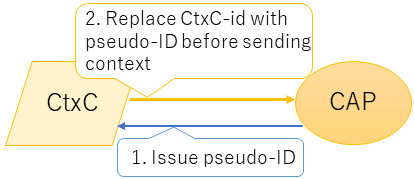}
	\caption{The case where CtxC is easily extensible}\label{fig:ctxc-easy}
\end{figure}
\textbf{When CtxC is easily extensible}
In this case, we assume CtxC is so extensible that it can share pseudonymous ID like web applications.
OpenID Connect(OIDC)\cite{oidc-core} and SAML\cite{saml-v2} are known as protocols for a share of pseudonymous IDs.

The way of linking contexts is explained in Fig. \ref{fig:ctxc-easy}.
First, CAP issues a pseudonymous identifier (pseudo-ID) to CtxC.
At this time, CAP stores the correspondence between the pseudo-ID and a CAP-id, and CtxC stores the correspondence between the pseudo-ID and a CtxC-id.
The issue of pseudo-IDs can be implemented, for example, using OpenID Connect ID tokens \cite{oidc-core}.
CtxC then replaces the CtxC-id in the context with the pseudo-ID and sends it to CAP.
This procedure allows the CAP to receive contexts with pseudo-IDs and to link contexts easily.
Also, in this procedure, CtxC and CAP are exchangeable so that CtxC may issue pseudo-ID.

\textbf{When CtxC is not easily extensible}
In this and subsequent sections, we will discuss methods of linking contexts in consideration of cases where CtxC is not easily extensible.
In this case, we assume that no additional implementation related to current implementation, such as embedded systems, can be made.
For CtxC and CAP to share a pseudo-ID, CtxC must authenticate the user via a browser or native application to map the pseudo-ID and CtxC-id.
Doing this would require additional implementation on CtxC's authentication for users or devices.
However, to need this additional implementation does not meet our assumption.

Therefore, in this study, we have designed the CtxC to implement additionally only a feature of transmitting context to CAP, which is independent of the existing CtxC's implementation.

\begin{figure}[htb]
	\centering
	\includegraphics[width=10cm]{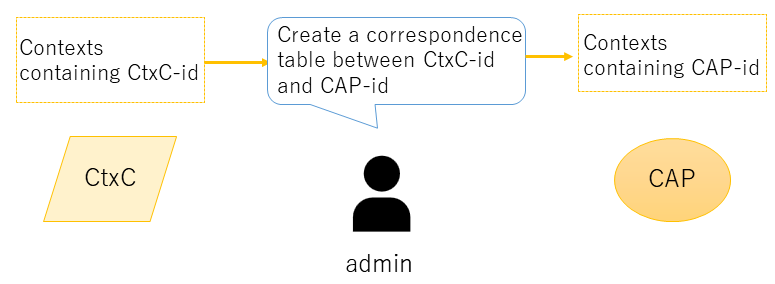}
	\caption{Administrator associates the CtxC-id with the CAP-id.}\label{fig:ctxc-admin}
\end{figure}

\textbf{When CtxC and CAP have the same administrator}\label{ctxc-admin}
As shown in Fig. \ref{fig:ctxc-admin}, in this case, the administrator knows the correspondence between a CtxC-id and a CAP-id so they can create the correspondence table in the CAP.
CAP uses this table to link contexts.

For example, suppose that a company operates CAP and that it would use an entrance control system using IC cards that CtxC manages by itself.
When registering an IC card at the time of joining the company, the administrator creates a correspondence between the IC card identifier and the employee ID and registers it in the CAP, so that the CAP can easily link the context.

With this approach, it is only necessary to transmit the context from CtxC to CAP in some way, and little additional implementation is required.
However, in this method, the CAP must trust that the administrator will maintain the correspondence table constantly.

\textbf{When CtxC authenticates using certificates}\label{ctxc-x509}
\begin{figure}[htb]
	\centering
	\includegraphics[width=7cm]{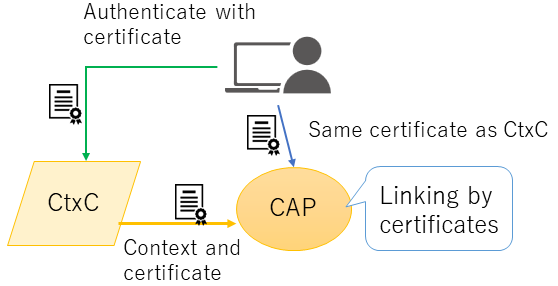}
	\caption{Overview of authentication by certificate}\label{fig:ctxc-cert}
\end{figure}

The method above requires implicit trust that the administrator will not do evil.
Therefore, we propose a method of linking contexts without implicitly trusting the administrator.
In this case, we assume the CtxC authenticates with a certificate and can send that part of the certificate and CtxC-id correspondence to the CAP with the context.
We also assume that the CA issues certificates correctly.
Examples of certificates available in this proposal are X.509 certificates and certificates stored in IC cards that control entering/exiting rooms.

In this assumption, CAP should verify the user/device has the private key of the certificate that the user/device uses for authentication in CtxC.
The method is illustrated in Fig. \ref{fig:ctxc-cert}.

We provide an overview in Fig. \ref{fig:ctxc-cert}.
As shown by the green arrow, CtxC requests a certificate from the user/device and authenticates using it.
Then, CtxC sends the certificate and context to CAP.
The CAP then requests the same certificate from the user/device used in CtxC to verify that the user/device has the certificate's private key.
It also verifies that the certificate has not been modified by validating the certificate chain.
This procedure allows the CAP to link contexts corresponding to certificates.

In this design, little additional implementation in CtxC is required.
Only the feature of sending contexts to the CAP is needed.
Also, we explained that the context is sent from CtxC to CAP, our method is applicable for a case where CtxC prepares an API and CAP obtains the context from it.

\section{An Example of CtxC and CAP implementation}
As example implementations to collect context from CtxC, this section shows how CAP links context with the certificates, using 802.1X\cite{1x} RADIUS server and a MDM service.

In this scenario, we consider the RADIUS server and MDM as CtxC.
The RADIUS server can collect communication logs in LAN and MDM can collect devices' states.
The communication logs have the information such as which access point the device connects, and are useful to locate the device.
Also, the devices' states have the information such as whether the device is being used with no known vulnerabilities, such as the OS version.
The RP can use these pieces of information as context to control access precisely.

We implemented the method of using certificates.
In our implementation, the Radius server authenticates devices using 802.1X EAP-TLS\cite{eaptls-rfc5216} and the MDM manages device certificates.
These satisfy the assumptions as CtxCs in that method.
Each CtxC sends a correspondence between context and certificate to the CAP, and the CAP links contexts by the certificate.

\begin{figure}[htb]
	\centering
	\includegraphics[width=8cm]{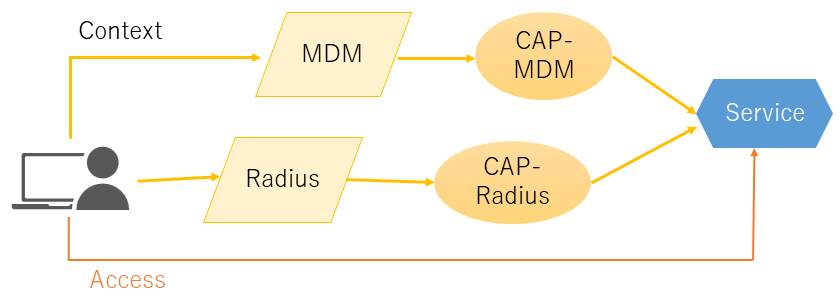}
	\caption{Overview of implement}\label{fig:zentaizou}
\end{figure}
Fig. \ref{fig:zentaizou} shows an overview of our implementation.
As CtxCs, the Radius server sends communication logs to CAP-Radius and the MDM sends devices' states to CAP-MDM.
They also send certificates with the context to CAPs.
The CAP then requests the device with certificate used in CtxC to link the context.
Afterward, the CAP associates the certificate with the CAP-id of the device.
These procedures allow the CAP to provide the RP with the context obtained from CtxC.

In this implementation, we used the FreeRADIUS server as CtxC.
Also, we used a Wi-Fi access point in our laboratory as an 802.1X authenticator.
This Wi-Fi access point uses EAP-TLS as an authentication protocol and uses a RADIUS server as an authentication server.
For the CAP implementation, we used the Go and Echo, a web framework for Go.

We monitored and sent two files, one is the RADIUS authentication log, and the other is the accounting log.
We have used fields of TLS-Client-Cert-Serial and TLS-Client-Cert-Issuer in the authentication logs.
The CAP can authenticate the device and link the context using these pieces of information.
From the accounting logs, we have used the field of Acct-Status-Type, which has the change of device's connectivity status as context.
We also have used the fields of Acct-Input-Octets and Acct-Output-Octets, which mean the device's traffic.
We implemented CAP to turn these contexts into useful states such as whether the device connects now.

We designed the CAP to gain context by periodically accessing Microsoft Intune's Managed Device API\footnote{https://docs.microsoft.com/ja-jp/graph/api/resources/intune-devices-manageddevice?view=graph-rest-beta}.
Intune is Microsoft's MDM service.

We have used the fields of osVersion, complianceState, lostModeState, and jailBroken to confirm that the device is securely maintained.
For example, we can use the OS version to know if the device is known as a non-compromised OS.
We implemented the ZTF using this information as a context.

\section{Concluding Remarks}
Zero Trust Federation (ZTF)\cite{hatakeyama-ztf} is a framework for authentication and authorization in ZTA under IdF.
To clarify the relationship between CAP and the data source, we separate CAP's roles into the collection and provision of context, and transfer the collection role to other entity we define as CtxC.
This separate leaves the CAP's role only to provide context.
We clarifies that the problem for CAP to obtain the context from CtxC is that the CAP must obtain a correspondence between the CtxC-id and CAP-id.
It is difficult to achieve this without making any assumptions.
Therefore, we addressed this problem in three cases based on use cases: when additional implementation is easy like web applications, when the administrators of the CtxC and the CAP are the same like recorders of enter/exit the rooms, and when the CtxC authenticates with a certificate like Radius servers using EAP-TLS.
Furthermore, we implemented the proposed method for the case where the RADIUS server and Intune are CtxC.
This implementation specifically shows the method of linking contexts when CtxC uses certificates for authentication.

The only context available by the proposed method is for cases where CtxC satisfies certain assumptions.
The availability of more diverse contexts is essential for making precise authorization decisions to more robustly protect resources.
Therefore, in future work, further methods should be devised to allow RPs to use CtxC contexts with a wider range of conditions.

\bibliographystyle{splncs04}
\bibliography{thesis}

\end{document}